\documentclass[twocolumn]{aastex62}

\def\lled{$L/L_{\rm Edd}$}
\def\rled{L/L_{\rm Edd}}

\def\lema{L/M_{\rm BH}}

\def\name{OGLE~J015531$-$752807}
\def\nme2{OGLE~J005907$-$645016}
\def\cand{31}
\def\cnd{33} 

\shorttitle{Discovery of two quasars at $z=5$ from the OGLE Survey}
\shortauthors{Koz{\l}owski et al.}

\begin{document}

\title{Discovery of two quasars at $z=5$ from the OGLE Survey}

\author[0000-0003-4084-880X]{Szymon~Koz{\l}owski}
\affiliation{Warsaw University Observatory, Al. Ujazdowskie 4, 00-478 Warszawa, Poland}
\affiliation{The OGLE Collaboration}
\email{simkoz@astrouw.edu.pl}

\author[0000-0002-2931-7824]{Eduardo Ba{\~n}ados}
\affiliation{The Observatories of the Carnegie Institution for Science, 813 Santa Barbara Street, Pasadena, California 91101, USA}

\author{A. Udalski}
\affiliation{Warsaw University Observatory, Al. Ujazdowskie 4, 00-478 Warszawa, Poland}
\affiliation{The OGLE Collaboration}

\author[0000-0003-2535-3091]{N. Morrell}
\affiliation{Carnegie Observatories, Las Campanas Observatory, 601 Casilla, La Serena, Chile}

\author[0000-0002-4863-8842]{A. P. Ji}
\affiliation{Hubble Fellow}
\affiliation{The Observatories of the Carnegie Institution for Science, 813 Santa Barbara Street, Pasadena, California 91101, USA}

\author[0000-0002-9658-6151]{{\L}. Wyrzykowski}
\affiliation{Warsaw University Observatory, Al. Ujazdowskie 4, 00-478 Warszawa, Poland}
\affiliation{The OGLE Collaboration}

\author[0000-0001-5990-6243]{A. Rau}
\affiliation{Max-Planck Institute for Extraterrestrial Physics, Giessenbachstr. 1, D-85748 Garching, Germany}

\author{P. Mr{\'o}z}
\affiliation{Warsaw University Observatory, Al. Ujazdowskie 4, 00-478 Warszawa, Poland}
\affiliation{The OGLE Collaboration}

\author{J. Greiner}
\affiliation{Max-Planck Institute for Extraterrestrial Physics, Giessenbachstr. 1, D-85748 Garching, Germany}

\author{M. Gromadzki}
\affiliation{Warsaw University Observatory, Al. Ujazdowskie 4, 00-478 Warszawa, Poland}

\author{M. K. Szyma{\'n}ski}
\affiliation{Warsaw University Observatory, Al. Ujazdowskie 4, 00-478 Warszawa, Poland}
\affiliation{The OGLE Collaboration}

\author{I. Soszy{\'n}ski}
\affiliation{Warsaw University Observatory, Al. Ujazdowskie 4, 00-478 Warszawa, Poland}
\affiliation{The OGLE Collaboration}

\author[0000-0002-9245-6368]{R. Poleski}
\affiliation{The Ohio State University, Department of Astronomy, 140 West 18th Avenue, Columbus, OH 43210, USA}
\affiliation{Warsaw University Observatory, Al. Ujazdowskie 4, 00-478 Warszawa, Poland}
\affiliation{The OGLE Collaboration}

\author[0000-0002-2339-5899]{P. Pietrukowicz}
\affiliation{Warsaw University Observatory, Al. Ujazdowskie 4, 00-478 Warszawa, Poland}
\affiliation{The OGLE Collaboration}

\author[0000-0002-2335-1730]{J. Skowron}
\affiliation{Warsaw University Observatory, Al. Ujazdowskie 4, 00-478 Warszawa, Poland}
\affiliation{The OGLE Collaboration}

\author{D. M. Skowron}
\affiliation{Warsaw University Observatory, Al. Ujazdowskie 4, 00-478 Warszawa, Poland}
\affiliation{The OGLE Collaboration}

\author{K. Ulaczyk}
\affiliation{Department of Physics, University of Warwick, Gibbet Hill Road, Coventry CV4 7AL, UK}
\affiliation{Warsaw University Observatory, Al. Ujazdowskie 4, 00-478 Warszawa, Poland}
\affiliation{The OGLE Collaboration}

\author{K. Rybicki}
\affiliation{Warsaw University Observatory, Al. Ujazdowskie 4, 00-478 Warszawa, Poland}
\affiliation{The OGLE Collaboration}

\author{P. Iwanek}
\affiliation{Warsaw University Observatory, Al. Ujazdowskie 4, 00-478 Warszawa, Poland}
\affiliation{The OGLE Collaboration}


\begin{abstract} 
We have used deep Optical Gravitational Lensing Experiment (OGLE-IV) images ($V \lesssim 23$ mag, $I \lesssim 23$ mag at $3\sigma$) of the Magellanic System, encompassing an area of $\sim$670 deg$^2$, to perform a search for high-$z$ quasar candidates.
We combined the optical OGLE data with the mid-IR Wide-field Infrared Survey Explorer (WISE) 3.4/4.6/12 $\mu$m data, and devised a multi-color selection procedure. 
We have identified \cnd\ promising sources and then spectroscopically observed the two most variable ones. 
We report the discovery of two high-$z$ quasars, \name\ at a redshift $z=5.09$ and  \nme2\ at a redshift of $z=4.98$.
The variability amplitude of both quasars at the rest-frame wavelength $\sim$1300\AA\ is much larger ($\sim$0.4 mag) than other quasars ($<0.15$ mag) 
at the same rest-frame wavelength but lower redshifts ($2<z<4$). To verify if there exist an increased variability amplitude in high-$z$
population of quasars, simply a larger sample of such sources with a decade long (or longer) light curves is necessary, which will be enabled by 
the Large Synoptic Survey Telescope (LSST) providing light curves for sources 3--4 mag fainter than OGLE.
\end{abstract}



\section{Introduction}

Being the brightest sources of continuous light in the Universe, quasars
serve as tracers of properties of the distant Universe. In particular,
they act as probes of the phase transition from the neutral to ionized Universe, the re-ionization era, at $6<z<11$ 
(e.g., \citealt{2001ApJ...560L...5D,2002AJ....123.1247F,2014A&A...571A...1P,2017MNRAS.466.4239G}). 
High redshift quasars must be inherently connected to
the cosmic structure in the early Universe, reflect their evolution, 
and also shed light upon (and influence) the evolution of the intergalactic medium.
It has been proposed that they may be used as ``standardizable candles''
to measure distances to the highest accessible redshifts (e.g., \citealt{1983ApJ...272...11P,2011ApJ...740L..49W,2013A&A...556A..97C,2015ApJ...815...33R}).
High-$z$ quasars could also shed light on the puzzle of 
the origin and growth of super-massive black holes (SMBHs) shortly after the Big Bang.
For example, \cite{2018Natur.553..473B} recently reported a quasar hosting a black hole with a mass of $8\times 10^8$ M$_\odot$ at a redshift of $z=7.54$,
just 700 million years after the Big Bang. The explanation of its origin and growth challenges some of the current SMBH formation models.
It requires a seed BH at $z \gtrsim 40$ with $100\lesssim M_{\rm BH0} \lesssim 10000$ M$_\odot$ to start with, the Eddington-limited accretion $L/L_{\rm Edd}=1$, and about 700 million years to grow such a BH, using the SMBH growth equation $t\approx0.043\ln{(M_{\rm BH}/M_{\rm BH0})}$ Gyr (e.g., \citealt{2005ApJ...620...59S,2014ApJ...784L..38M,2016ApJ...828...26M,2018Natur.553..473B}).

For all the above reasons, high redshift quasars have been and still are the target of intensive searches,
providing to date approximately 320 quasars at $z>5$, 90 at $z>6$ 
(e.g., \citealt{2001AJ....122.2833F,2013ApJ...779...24V,2016ApJS..227...11B,2016ApJ...819...24W,2018PASJ...70S..35M}) and two at $z>7$ (\citealt{2011Natur.474..616M,2018Natur.553..473B}).

In general, quasar spectra or spectral energy distributions (SEDs)
are well understood. In the UV and optical, the SED is dominated by emission from an accretion disk, forming a ``blue'' continuum 
up to about 1$\mu$m (rest-frame). At longer wavelengths, up to mid/far-IR, the quasar emission is dominated by a dusty torus, forming
an ``IR bump''. On top of these features, there exists a number of known and generally understood broad and narrow emission lines, of which the most common are Ly$\alpha$, CIV, CIII], MgII, H$\beta$, [OIII], and H$\alpha$. 

A typical method for selecting high-$z$ quasars is
to observe them in a number of UV-optical-IR filters and search for a sudden drop in luminosity below a certain wavelength. 
Such a break appears at wavelengths shorter than Ly$\alpha$ at 1216\AA\ (rest-frame) for high-$z$ quasars due to a significant absorption by 
clumps of neutral hydrogen along the line of sight. For example, an $i$-band (7625\AA; \citealt{1996AJ....111.1748F}) ``dropout'' simply means that a quasar is not detectable in the $i$-band (and other shorter wavelength bands) but is certainly detectable in redder filters such as the $z$-band (9134\AA). We can quickly infer the redshift in such a case that must be $z \approx 6$ (e.g., $\sim$8500\AA/1215.67${\rm \AA}=1+z$).

Another powerful method of selecting quasars is the mid-IR selection (e.g., \citealt{2004ApJS..154..166L,2005ApJ...631..163S,2012ApJ...753...30S,2013ApJ...772...26A}). The dusty torus absorbs a fair fraction of radiation coming from the disk and re-emits it in IR, what makes AGNs readily visible, bright sources
in mid-IR. Another obvious step is to combine both the optical and IR selection methods (e.g., \citealt{2009ApJ...701..508K,2016ApJ...829...33Y}),
and also supplementing them with other wavelengths (X-ray, radio; e.g., \citealt{2008ApJ...679.1040G,2015ApJ...804..118B}). AGN are also known as variable sources (e.g., \citealt{2004ApJ...601..692V,2010ApJ...721.1014M}) with amplitudes of about 0.25 mag in optical bands, provided the light curves are several-years-long in rest-frame (e.g, \citealt{2016ApJ...826..118K}).

In this paper, we combine the optical ``$V$-band dropout'' method with the mid-IR quasar selection methods, 
where we pay close attention to AGN candidates that are variable sources in $I$-band. Because there exists a well known anti-correlation
between the variability amplitude and the luminosity $L$ and/or Eddington ratio ($\rled \propto \lema$) for AGN (e.g., \citealt{2010ApJ...721.1014M,2016ApJ...826..118K}), their variability is expected to be tiny given the necessary high luminosity for high-$z$ quasars to be observable. 

The paper is organized as follows. In Section~\ref{sec:method}, we will present methods used to select high-$z$ quasar candidates, 
while in Section~\ref{sec:quasar}, we will describe the two high-$z$ confirmed quasars in the OGLE survey. In Section~\ref{sec:other}, we will present the remaining candidates. The paper is discussed in Section~\ref{sec:summary}.


\section{The method}
\label{sec:method}

The OGLE-IV survey (\citealt{2015AcA....65....1U}) has monitored the Magellanic Clouds area since 2010. Each 1.4 sq. deg. field out of 560 has been observed
at least 100 times in the $I$-band. We have stacked typically 80 and 20 high-quality, good-seeing, low-background images in $I$-band and $V$-band (Udalski in prep.), respectively, to increase the depth of detection to about $I<23.0$ mag (3$\sigma$) and $V<23.0$ mag (3$\sigma$), both in the Vega magnitude system.

\begin{figure*}
\centering
\includegraphics[width=6.9cm]{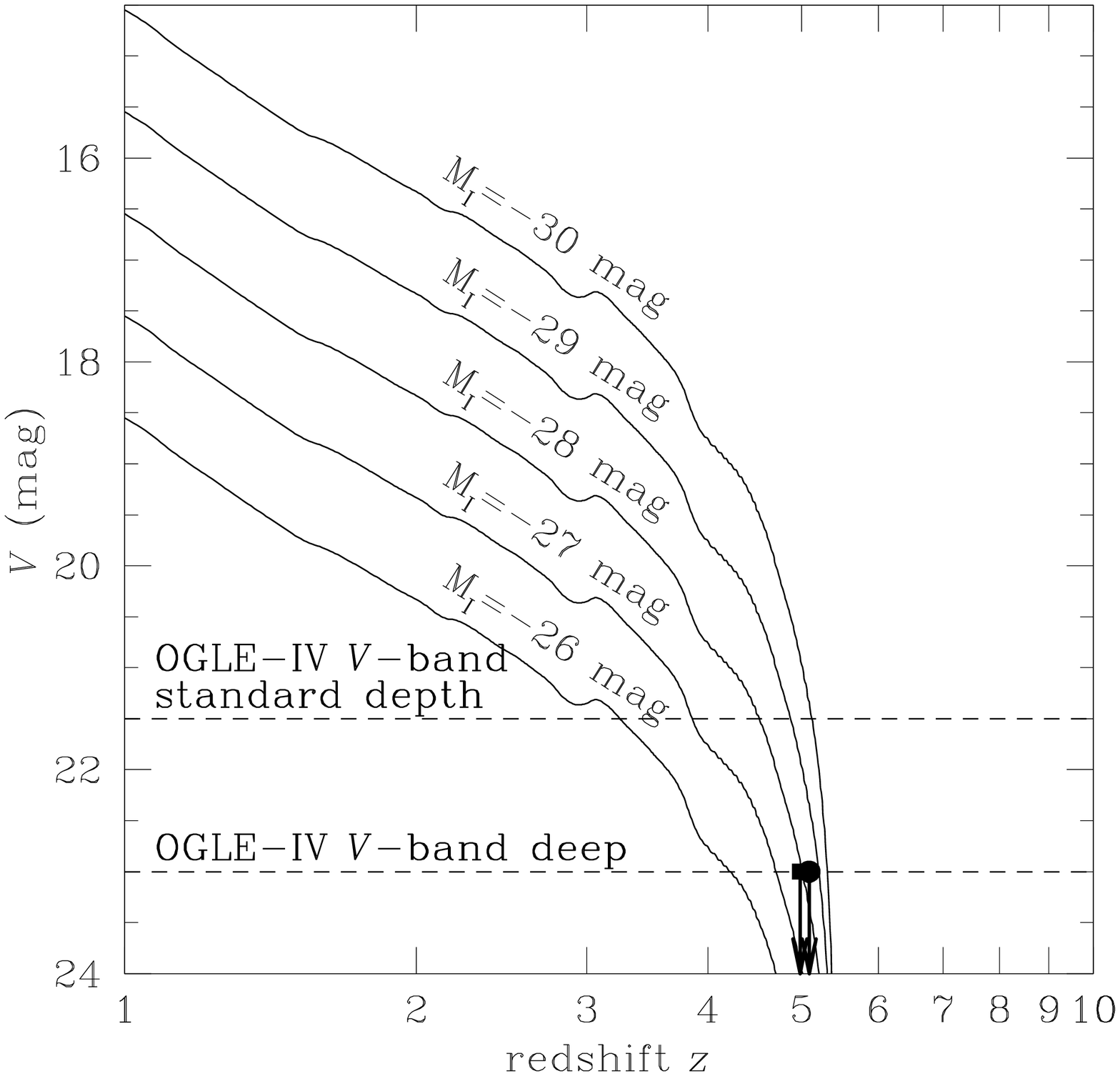}\hspace{0.5cm}
\includegraphics[width=6.9cm]{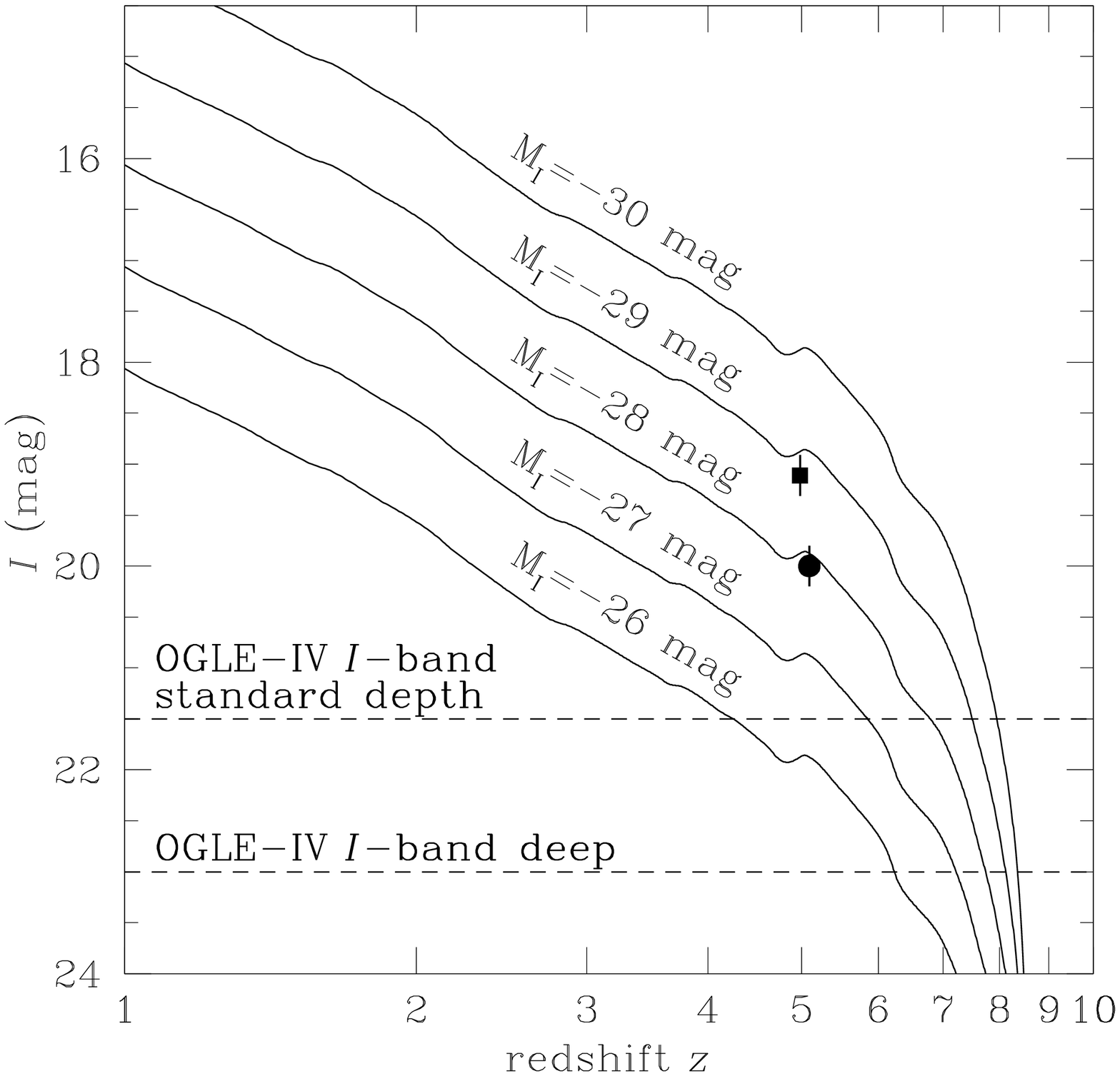}\\
\caption{Synthetic observed magnitudes for quasars in $V$-band (left panel) and $I$-band (right) OGLE filters as a function of redshift and the absolute $I$-band magnitude in a standard $\Lambda$CDM Universe with $\Omega_M=0.29$, $\Omega_\Lambda=0.71$, and $H_0=70$ km s$^{-1}$ Mpc$^{-1}$, and calculated using the 
mean SDSS quasar spectrum from \cite{2001AJ....122..549V}.
The two quasars detected in OGLE, \name\ at $z=5.09$ and \nme2 at $z=4.98$ are marked in both panels as a large dot and a square, respectively.}
\label{fig:colors}
\end{figure*}

In Figure~\ref{fig:colors} we present the expected observed $V$- and $I$-band magnitude of quasars as a function of redshift and absolute $I$-band magnitude.
We used a standard $\Lambda$CDM model with $\Omega_M=0.29$, $\Omega_\Lambda=0.71$, $H_0=70$ km s$^{-1}$ Mpc$^{-1}$, and the \cite{2001AJ....122..549V}
spectrum to calculate k-corrections. For redshifts $z>4$, the Ly$\alpha$ line is redshifted out of the $V$-band filter, while the continuum flux is highly attenuated by neutral hydrogen in the Universe (a leftover from the re-ionization era), leading to a sudden drop of luminosity in $V$-band. This is why quasars at $z\gtrsim4$ become undetectable in the OGLE $V$-band (the left panel of Figure~\ref{fig:colors}). At the same time, the $I$-band filter is dominated by the maximum flux from an accretion disk for $z \lesssim 6$. The Ly$\alpha$ line is redshifted out of the $I$-band filter at about $z\gtrsim6.5$ leading to a sudden drop in luminosity above this redshift (the right panel of Figure~\ref{fig:colors}). The OGLE-IV filter setup is most sensitive to the $V$-band dropout method with the target redshifts of $4\lesssim z \lesssim 6$.

\begin{figure*}
\centering
\includegraphics[width=7.5cm]{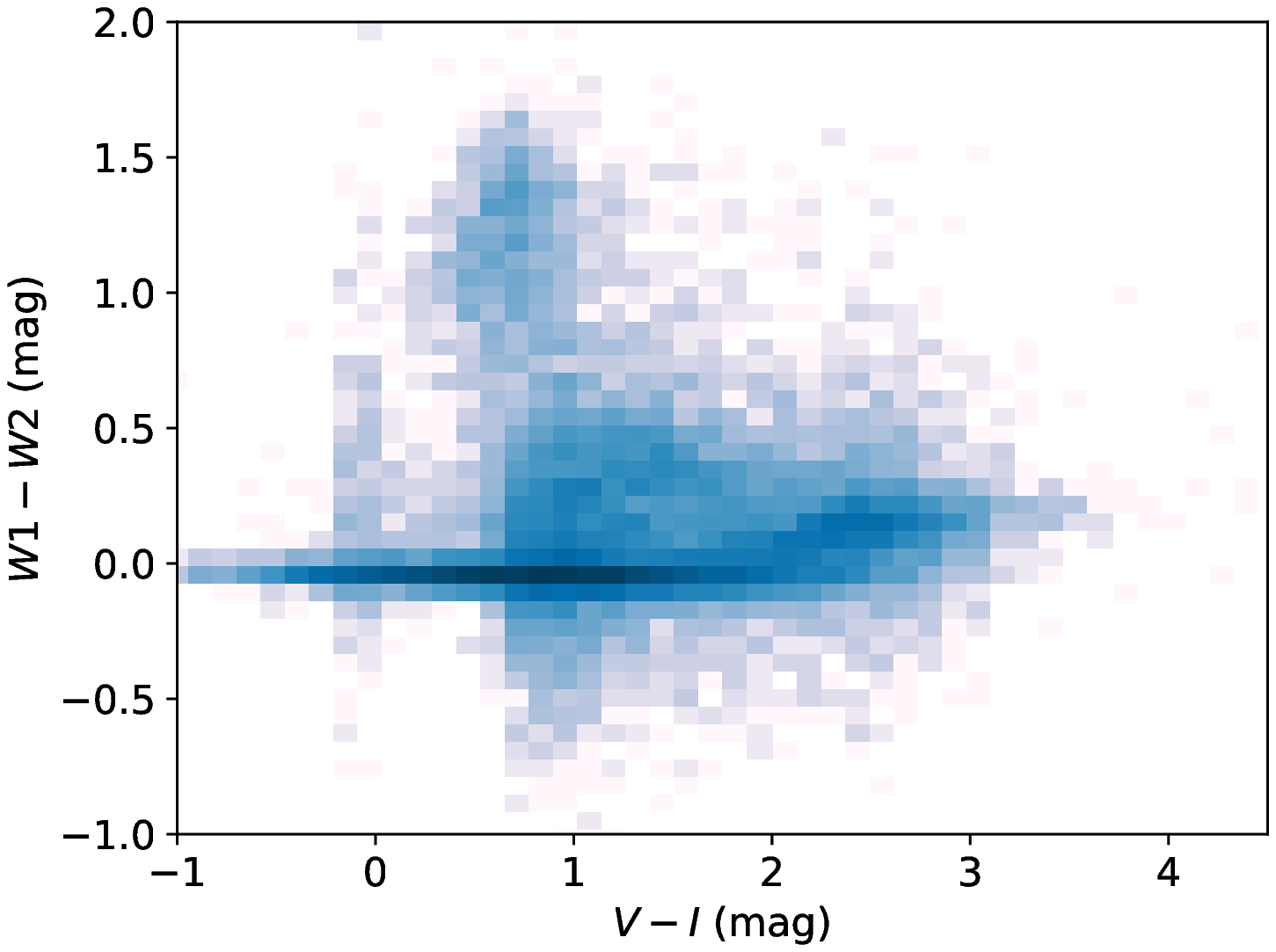}
\includegraphics[width=7.5cm]{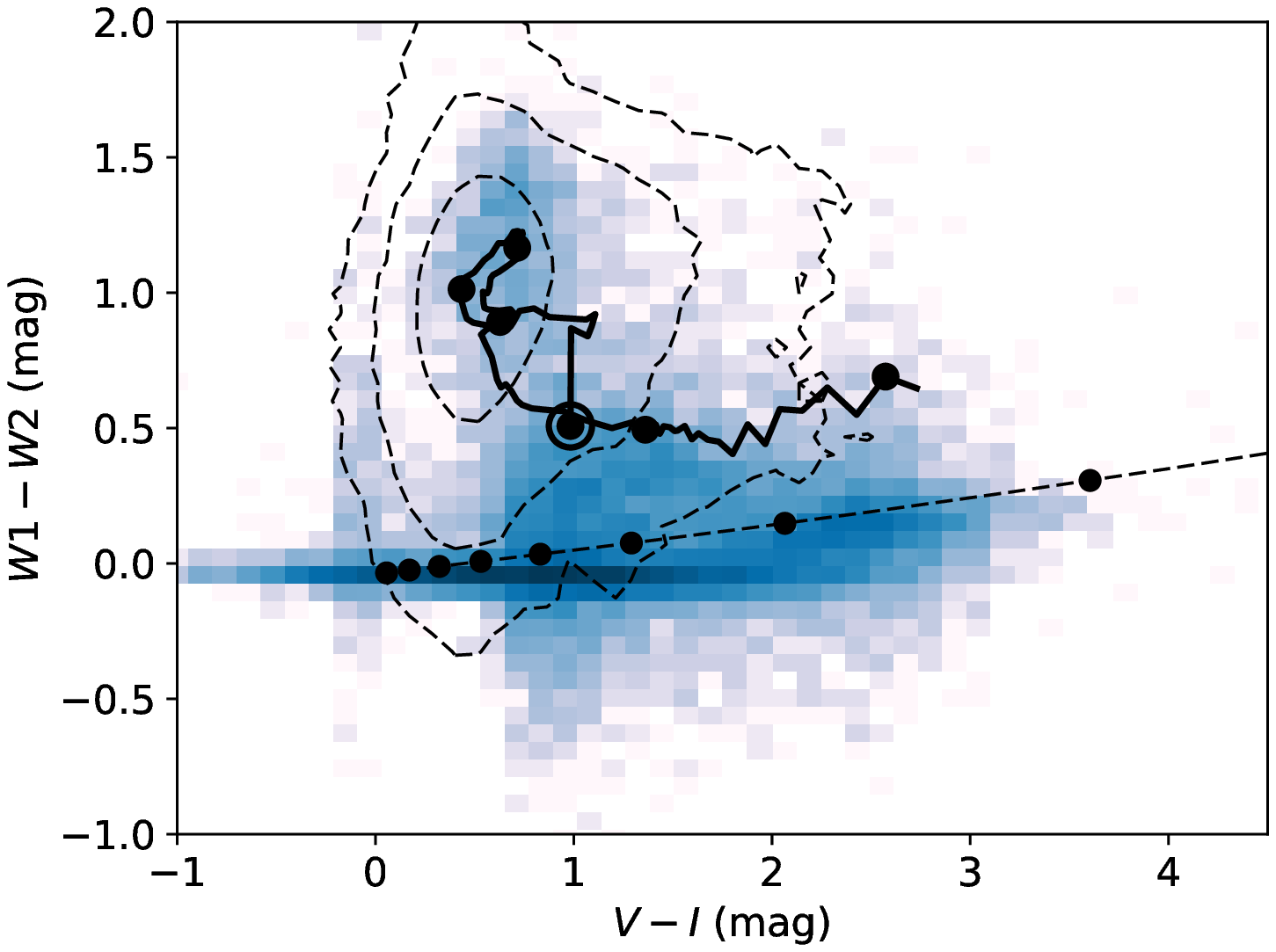}\\
\includegraphics[width=7.5cm]{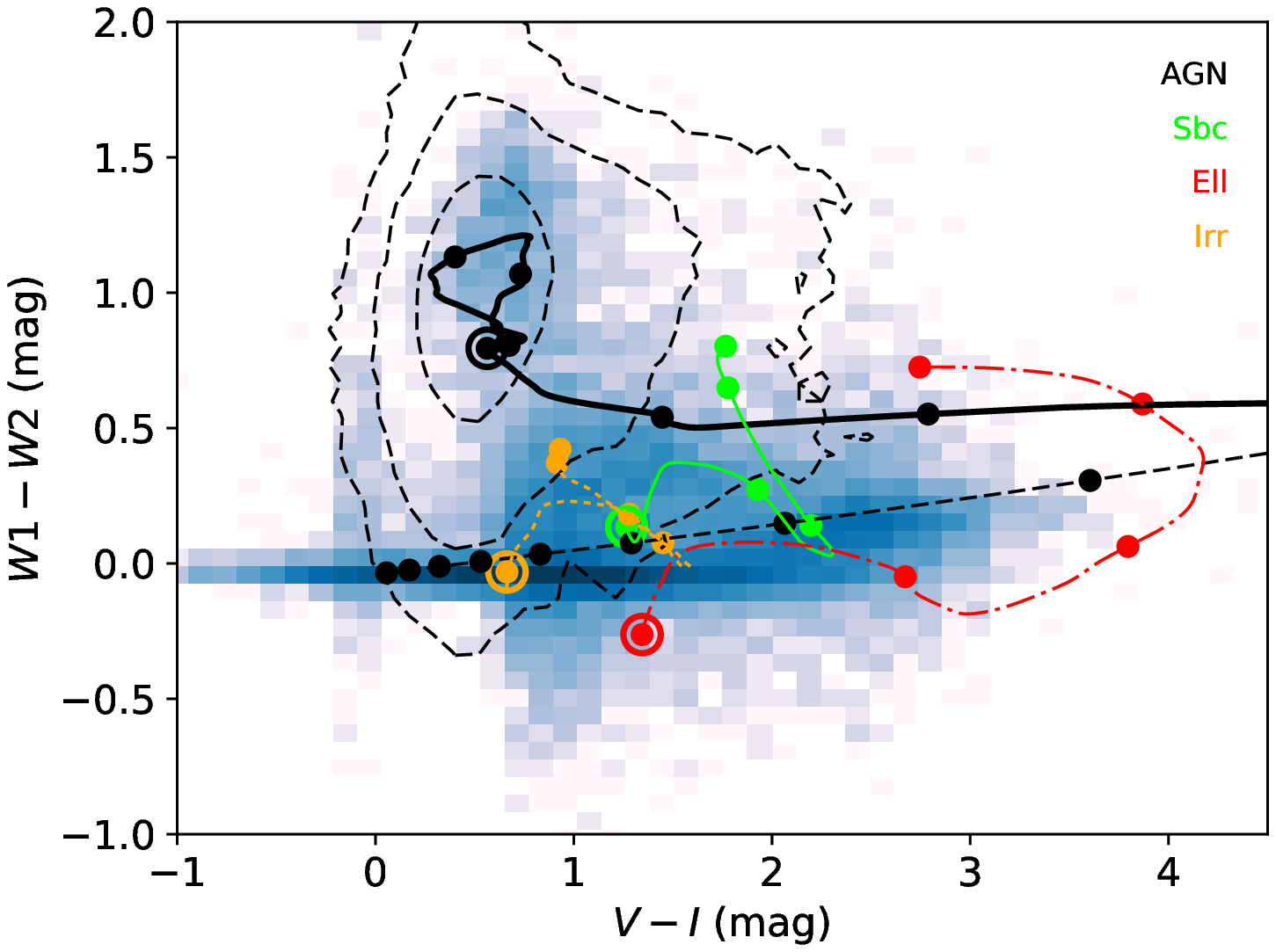}
\includegraphics[width=7.5cm]{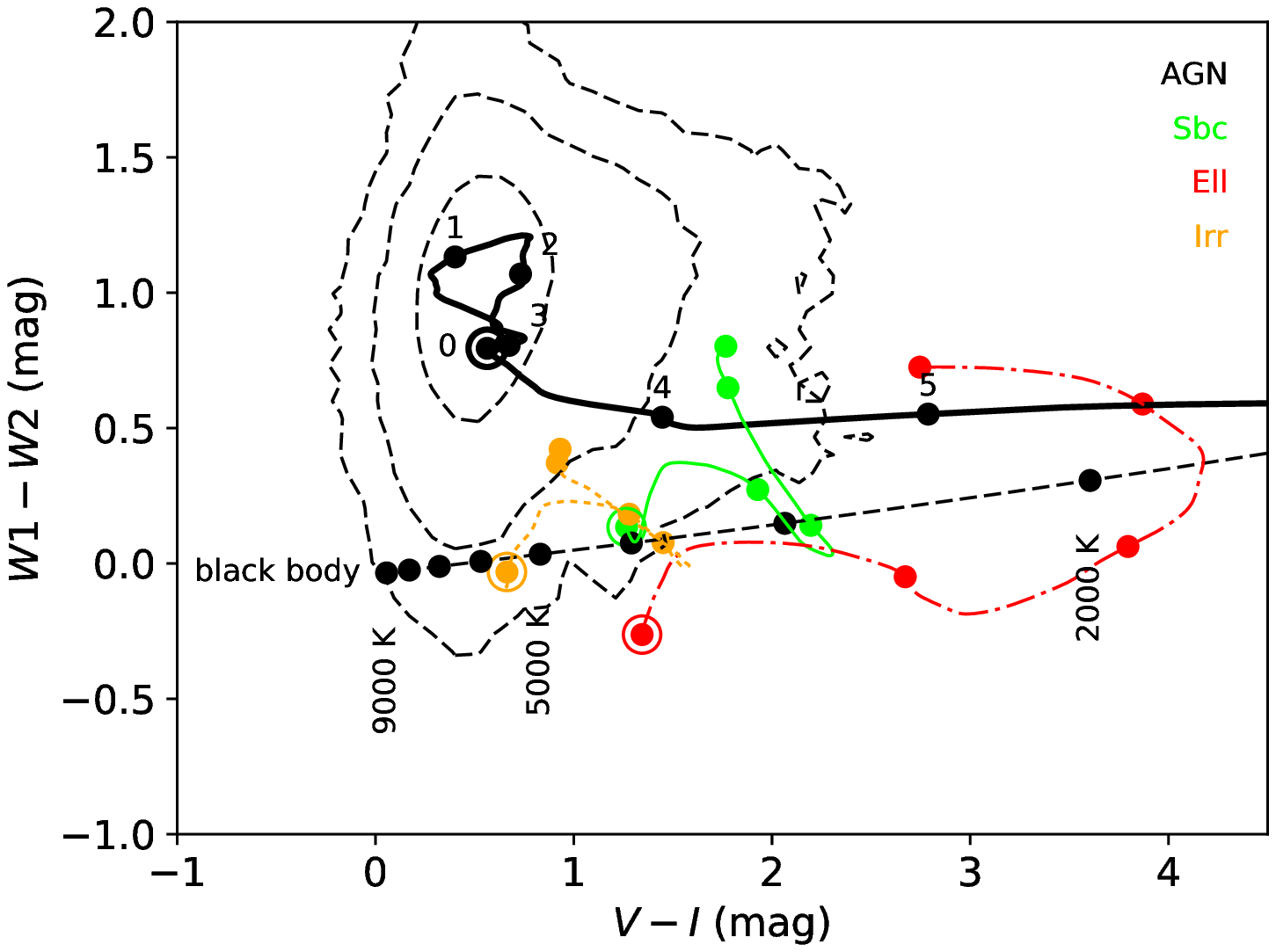}
\caption{OGLE--WISE colors and density maps for sources in the 30 sq. deg. area of the Magellanic Bridge. 
{\it Top-left:} The density map is shown.
{\it Top-right:} The density map along with the AGN density contours (the inner one with 1000, middle 100, and outer 10 AGN/bin) is shown. The solid line shows the color-color AGN track based on SDSS $ugriz$ magnitudes converted to OGLE $VI$ system as a function
of redshift, where the bulls-eye is $z=0$ and each dot marks a redshift increment of 1. The dashed, near horizontal, line is the color-color track
for a black body (a simplified model for stars) as a function of temperature, where the leftmost dot marks the temperature of 9000K, while the other dots show a temperature decrement by 1000K (going from left to right).
{\it Bottom-left:} The solid line shows the color-color AGN track obtained from a synthetic AGN spectrum as a function
of redshift, where the bulls-eye is $z=0$, and each dots mark a redshift increment of 1. We also show color-color tracks for spiral (solid green line), elliptical (red, dash-dot), and irregular (dotted orange) galaxies obtained from the \cite{2010ApJ...713..970A} templates. Again, the bulls-eye is for $z=0$, while each dot marks a redshift increment of $0.5$.
{\it Bottom-right:} We remove the density map for better visibility of the tracks and contours. In-depth discussion of the figure is provided in Section~\ref{sec:method}.}
\label{fig:colors_ugriz_IR}
\end{figure*}

We have matched the deep OGLE $VI$ photometric maps to the AllWISE infrared data, where 
AllWISE\footnote{\tt http://wise2.ipac.caltech.edu/docs/release/allwise/} is a merger of WISE 
(\citealt{2010AJ....140.1868W}) and NEOWISE (\citealt{2011ApJ...731...53M}) projects.
In the top-left panel of Figure~\ref{fig:colors_ugriz_IR} we present the $W1-W2$ ($3.4\mu$m$-4.6\mu$m) and $V-I$ color density map for a selected 30 sq. deg. area in between the Magellanic Clouds.
The extinction towards this area is typically $A_V=0.09$ mag and $A_I=0.05$ mag (\citealt{2011ApJ...737..103S}), and for both $V-I$ and WISE colors were assumed to be negligible in this quasar search.

We used the twelfth quasar data release (DR12Q, \citealt{2017A&A...597A..79P}) of the Sloan Digital Sky Survey (SDSS) to locate quasars in the
optical--mid-IR color-color diagram. 
Since SDSS provides the photometric maps in $ugriz$ filters (\citealt{1996AJ....111.1748F}), we converted them into $VI$ maps. We used the combined 
UV-optical quasar spectrum from \cite{2001AJ....122..549V} and the low resolution spectral energy distribution from \cite{2010ApJ...713..970A}
to transform individual DR12Q SDSS quasar colors to the OGLE colors. In the top-right panel of Figure~\ref{fig:colors_ugriz_IR}, 
we present the color-color tracks for quasars as a function of redshift (solid line), where the bulls-eye corresponds to $z=0$, 
and each dot marks the increase of redshift by 1. On top of the color-coded density map, we also show contours (dashed lines) for over 190,000
AGNs from SDSS DR12Q matched to AllWISE (with $\sigma_{W1}<0.1$ mag, $\sigma_{W2}<0.1$ mag, and $\sigma_{W3}<0.2$ mag), where the three levels are 10, 100, and 1000 objects per bin (0.125 mag in $V-I$ and 0.06 mag in $W1-W2$). 

To understand the density shapes in the color-color map in the top-left panel of Figure~\ref{fig:colors_ugriz_IR}, we also calculated color-color tracks for spiral, elliptical, and irregular galaxies using the \cite{2010ApJ...713..970A} models, shown as color-coded dotted, dash-dotted, and solid lines, where again the bulls-eye is for objects at $z=0$ and the dots mark the increase of redshift by 0.5, up to $z=2.0$ (bottom panels of Figure~\ref{fig:colors_ugriz_IR}).
The dashed line presents the black body color (a simplified model for stars) as a function of temperature, where the most left hand side dot is for 9000K and the temperature drops by 1000K for each dot toward right (as the $V-I$ color increases). 

The quasar tracks (solid black line) in both bottom panels of Figure~\ref{fig:colors_ugriz_IR} are calculated 
from the synthetic SED (combined the \citealt{2001AJ....122..549V} and \citealt{2010ApJ...713..970A} data), 
in opposition to the color--color tracks in the top-right panel
where we converted the SDSS colors to the OGLE colors for individual objects. Both tracks show a high level of similarity.

Since we are interested in high-$z$ quasars it is clear we should search for objects with large $V-I$ colors (or with no detection in $V$, the $V$-band dropout) and the $(W1-W2)$ colors of about 0.5 mag. We used the following cuts to select high-$z$ quasar candidates:
a point source with $18.5<I<20.5$~mag (to exclude low-$z$ contaminating extended galaxies, and to be still able see the variability, albeit heavily obscured by the photometric noise),
$(V-I)>1.5$ mag (preferably with non-detection in V),
$W1>16$ mag (to avoid potentially bright IR sources from the Magellanic Clouds), 
$0.3<(W1-W2)<0.9$ mag,
$(W2-W3)>2.5$ mag (to avoid brown dwarfs with $(W2-W3)\lesssim2.5$ mag from \citealt{2011ApJS..197...19K}),
$(I-W1)<3.2$ mag. 
After these cuts we are left with 38 sources, but as we will show in the next section, we removed additional five sources due to high proper motions, finally leaving us with \cnd\ candidates.
Adding the requirement of visually detectable variability\footnote{The variability amplitude is higher than the photometric noise amplitude for sources with similar brightness, it is not periodic, it is not a white noise, but must be consistent with a stochastic-type variability.}, we were able to find three variable high-$z$ quasar candidates, of which as of now two have been studied spectroscopically (this paper).


\section{The high-$z$ quasars}
\label{sec:quasar}

\subsection{\name}
\name\ is our first variable high-$z$ quasar candidate to be followed-up. The source is located at (RA, Decl.$)=($01:55:31.08, $-75$:28:07.13); it has not been detected in the $V$-band ($V>23.0$~mag), while its $I$-band magnitude is $I \approx 20.0$~mag, with the peak-to-peak variability amplitude of $\sim$0.5 mag. It is also located in two overlapping OGLE-IV fields, where its ID is: MBR104.01.297 and MBR110.25.427.
We observed the candidate with GROND in $g'$, $r'$, $i'$, $z'$, $J$, $H$, $K$ filters (\citealt{2008PASP..120..405G}) on January 10, 2018,
on the MPG 2.2m telescope at the ESO La Silla Observatory, Chile (see Table~\ref{tab:mags}). 
The GROND  $g^\prime r^\prime i^\prime z^\prime$ data were calibrated with a short subsequent exposure of an equatorial field against cataloged
 magnitudes of field stars from the SDSS, and the $JHK_{\rm s}$ against the 2MASS catalog.

\begin{figure}
\centering
\includegraphics[width=8.2cm]{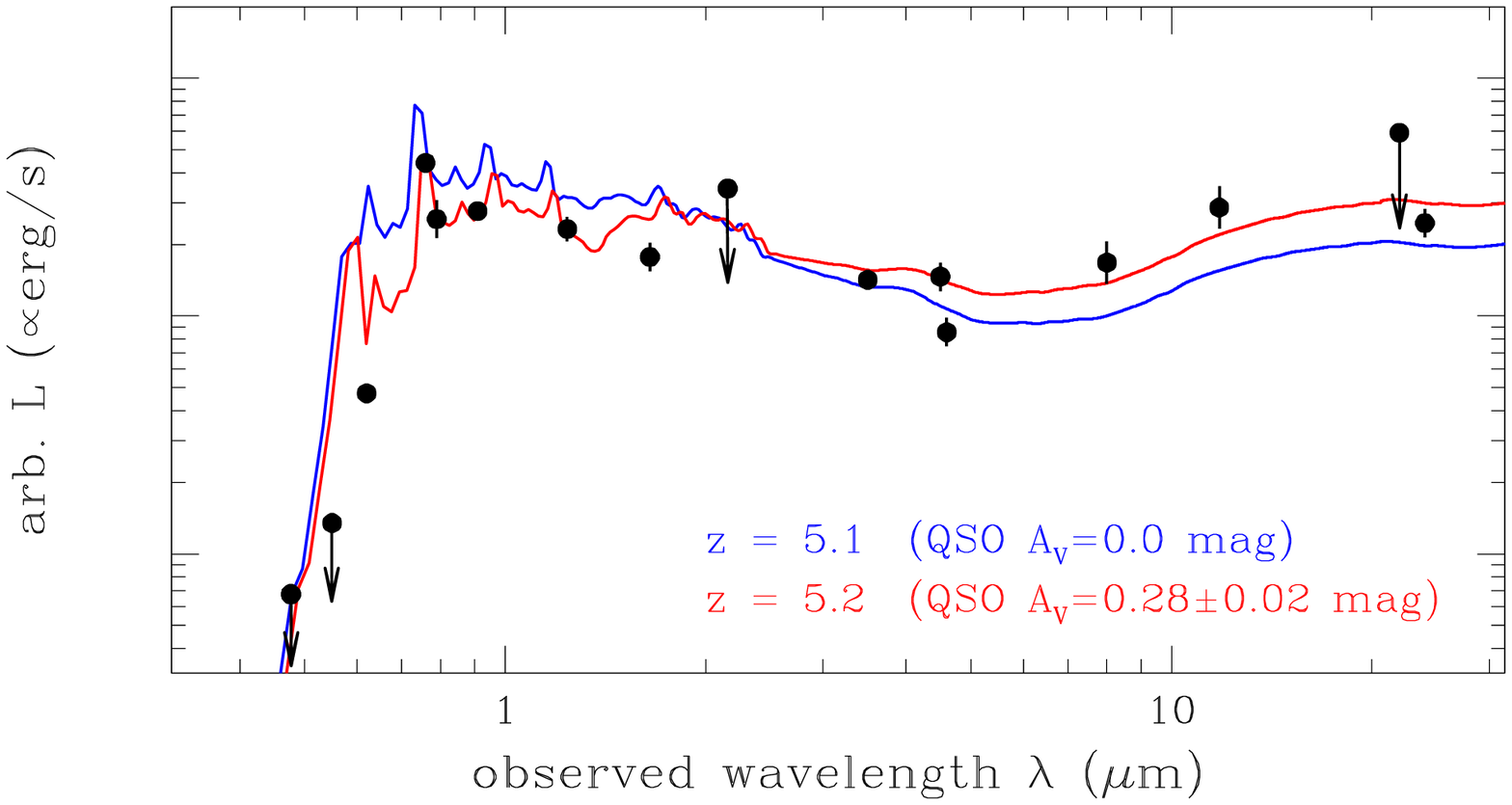}
\caption{Spectral energy distribution (SED) for \name\ is shown. Black dots are broad-band measurements from OGLE ($VI$), GROND ($g'r'i'z'JHK$), WISE (3.4, 4.6, 12, and 22$\mu$m) and Spitzer (4.5, 8.0, and 24 $\mu$m). Prior to the spectroscopic confirmation, we fitted the data using the combined \cite{2001AJ....122..549V} (UV-optical) and \cite{2010ApJ...713..970A} (IR) models with (red line) and without (blue line) the internal AGN host extinction (using the \cite{1989ApJ...345..245C} model and $R_V = 3.1$), obtaining the redshift estimate of $z\approx5.2$ and $z\approx5.1$, respectively.}
\label{fig:sed}
\end{figure}

The object is also detected in WISE 3.4 ($W1$), 4.6 ($W2$), 12 ($W3$) and we have an upper limit for the 22 ($W4$) $\mu$m. There are also available data from Spitzer SAGE-SMC IRAC 4.5 $\mu$m ($C2$) and 8.0 $\mu$m ($C4$) images, and also Spitzer MIPS 24$\mu$m ($C5$) image (\citealt{2007ApJ...655..212B,2011AJ....142..102G}). We converted these magnitudes into fluxes and modeled this SED 
using the combined \cite{2001AJ....122..549V} (UV-optical) and \cite{2010ApJ...713..970A} (IR) models with and without the internal AGN host extinction (using the \cite{1989ApJ...345..245C} model and $R_V = 3.1$), obtaining the redshift estimate of $z\approx5.2$ and $z\approx5.1$, respectively (Figure~\ref{fig:sed}).

\begin{deluxetable}{lcccc}
\centering
\tablecaption{Photometric measurements for \name.\label{tab:mags}}
\tablewidth{\columnwidth}
\tablehead{Filter & $\lambda$ ($\mu$m) &system & magnitude & uncert.}
\startdata
$g'$  & 0.48 & AB & $>24.9$ & $3\sigma$ \\
$V$   & 0.55 & Vega & $>23.0$ & $3\sigma$ \\
$r'$  & 0.62 & AB & 22.50 & 0.08 \\
$i'$  & 0.76 & AB & 19.86 & 0.08 \\
$I$   & 0.79 & Vega & $\sim$$20.0$ & $\cdots$ \\
$z'$  & 0.91 & AB & 20.17 & 0.09 \\
$J$   & 1.24 & AB & 20.03 & 0.14 \\
$H$   & 1.65 & AB & 20.00 & 0.15 \\
$K$   & 2.16 & AB & $>19.0$ & $3\sigma$ \\
$W1$  & 3.35  & Vega & 16.80 & 0.06 \\
$C2$  & 4.51  & Vega & 15.86 & 0.13 \\
$W2$  & 4.60  & Vega & 16.36 & 0.14 \\
$C4$  & 7.98  & Vega & 13.97 & 0.21 \\
$W3$  & 11.56 & Vega & 12.26 & 0.22 \\
$W4$  & 22.09 & Vega & $>9.29$ & $2\sigma$ \\
$C5$  & 23.68 & Vega & 10.00 & 0.13
\enddata
\tablecomments{The GROND data were corrected for the Galactic extinction with $E(B-V)=0.034$ mag. 
$C$- and $W$-filters are from Spitzer and WISE, respectively, and their central wavelength is given in the second column.}
\end{deluxetable}

\begin{figure}
\centering
\includegraphics[width=8.2cm]{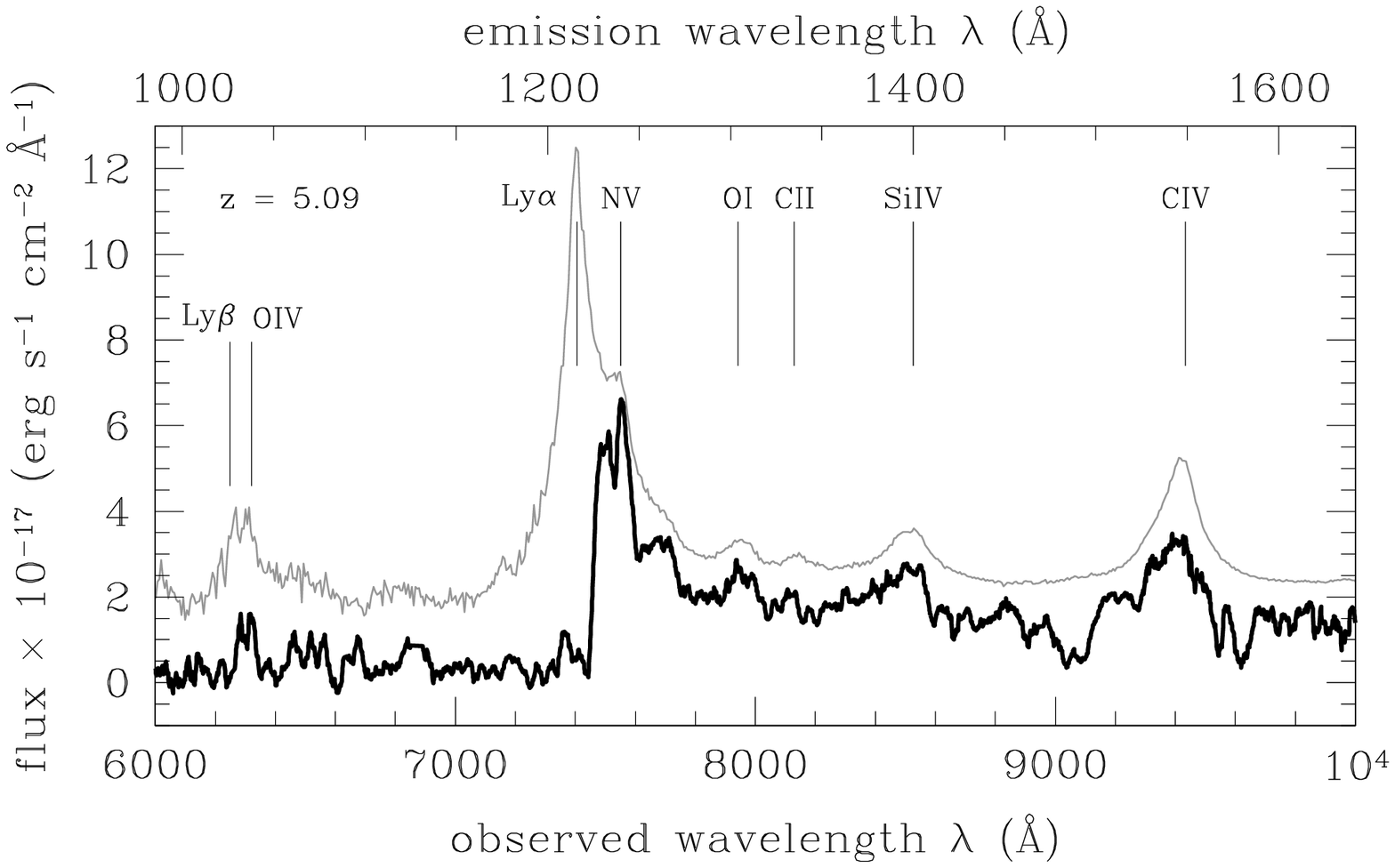}\vspace{0.3cm}
\includegraphics[width=8.2cm]{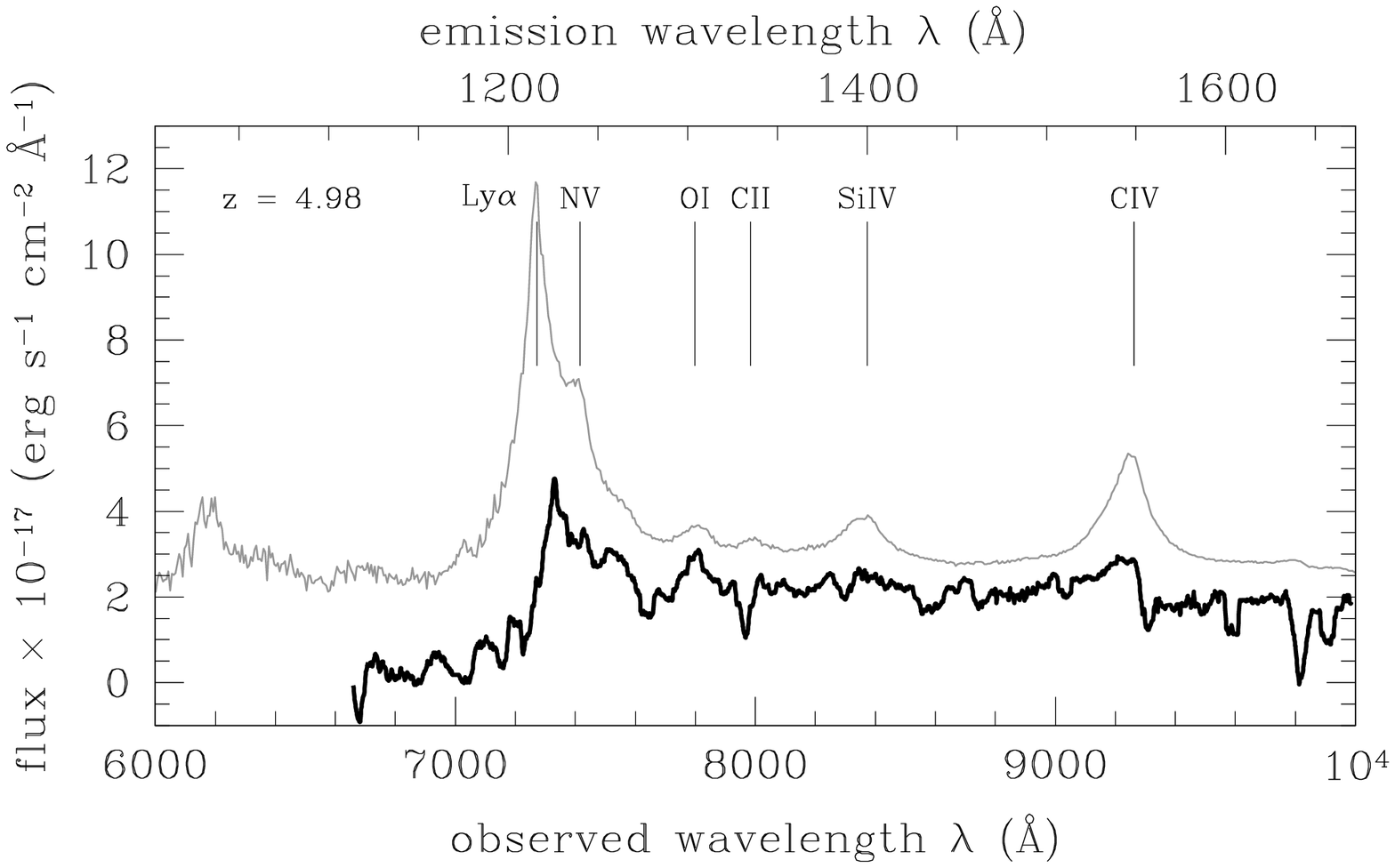}
\caption{The LDSS3 spectra of \name\ (top) and \nme2\ (bottom) are shown as black solid lines. For comparison, the SDSS AGN spectrum from \cite{2001AJ....122..549V} (gray) is shown, along with major AGN lines marked with vertical lines.}
\label{fig:spec}
\end{figure}

\name\ is confirmed as a quasar with a 900 seconds spectrum taken on 2018 January 25 using the Low-Dispersion Survey Spectrograph (LDSS3) at the Clay  (Magellan) Telescope at the Las Campanas Observatory, Chile. The observation was carried out in 1.2 arcsec seeing using the VPH-RED grism with the 1~arcsec wide ``center'' long-slit. The spectrum shows a sharp Ly$\alpha$ break at around 7500 \AA, indicating that it is a quasar at $z>5$. The spectrum was reduced using standard IRAF routines, including bias subtraction, flat fielding, sky subtraction, wavelength calibration, and spectrum extraction. We observed the standard star LTT1788 to perform the flux calibration (\citealt{1992PASP..104..533H,1994PASP..106..566H}). The final spectrum, shown in Figure~\ref{fig:spec}, was scaled to match the $i$-band GROND follow-up photometry to account for slit losses.

\begin{figure}
\centering
\includegraphics[width=8.2cm]{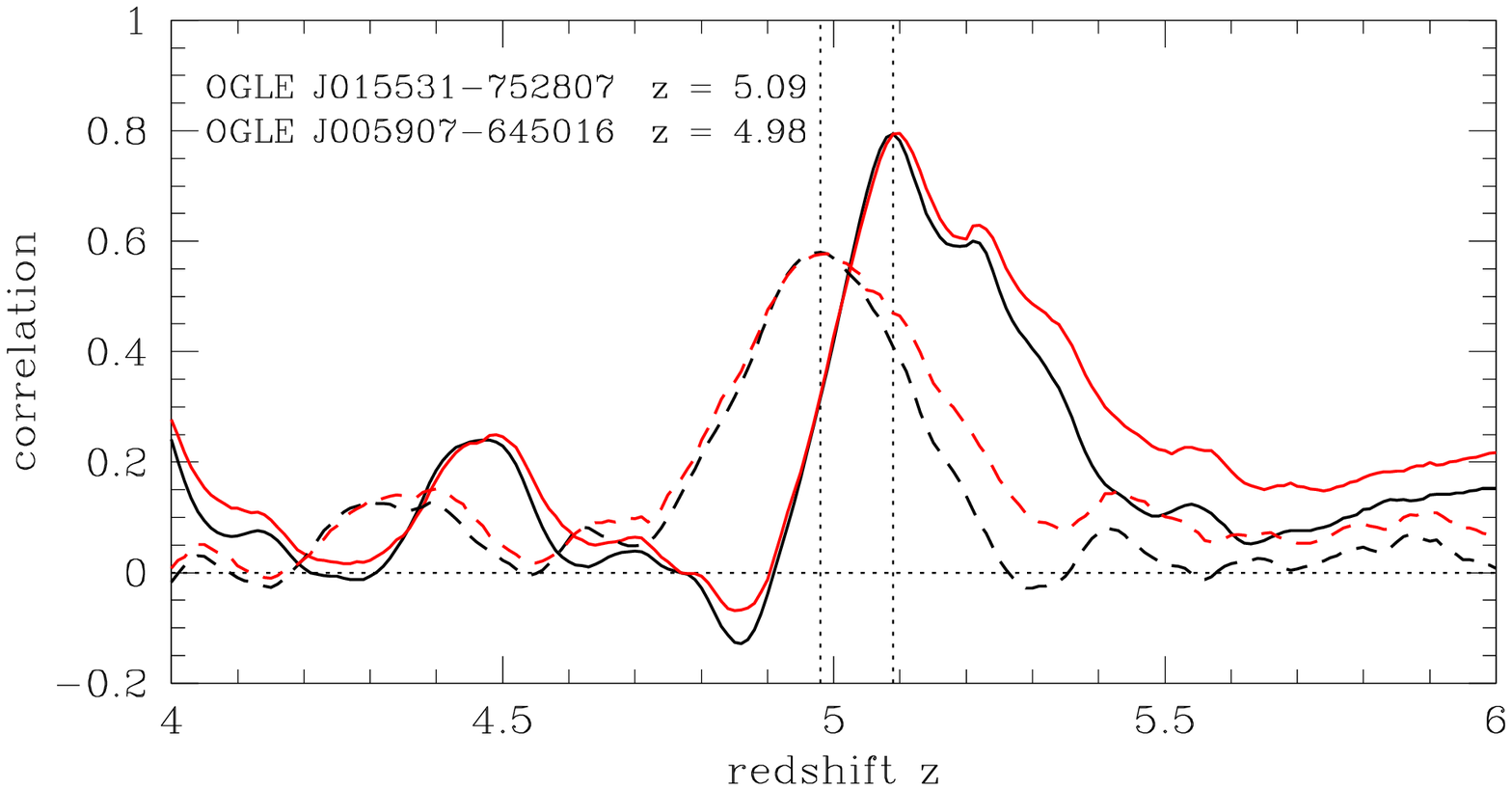}
\caption{Redshift estimates for \name\ and \nme2\ from correlating the LDSS3 spectrum (from Figure~\ref{fig:spec} for $\lambda_{\rm observed}>7500$\AA\ and $\lambda_{\rm observed}>7320$\AA, respectively) with the \cite{2001AJ....122..549V} SDSS AGN spectrum (black) and the \cite{2016A&A...585A..87S} bright QSO spectrum (red).
The correlation peaks at 0.8 (0.6) for $z=5.09$ ($z=4.98$) marked with the vertical dotted line for \name\ (\nme2). The horizontal dotted line marks the lack of correlation.}
\label{fig:zest}
\end{figure}

To find the redshift, we have cross-correlated the \name\ spectrum for $\lambda_{\rm observed}>7500$\AA\ with the \cite{2001AJ....122..549V} SDSS AGN spectrum and also with the \cite{2016A&A...585A..87S} bright QSO spectrum.
The Pearson correlation peaks at $r=0.8$ for $z=5.09$ (Figure~\ref{fig:zest}, solid lines). High-$z$ quasars are known to have the CIV line
strongly blueshifted with respect to other lower ionization lines such as MgII (e.g., \citealt{2011AJ....141..167R,2017ApJ...849...91M}), 
which may have an impact on our measurement of the redshift.

\begin{figure}
\centering
\includegraphics[width=8.2cm]{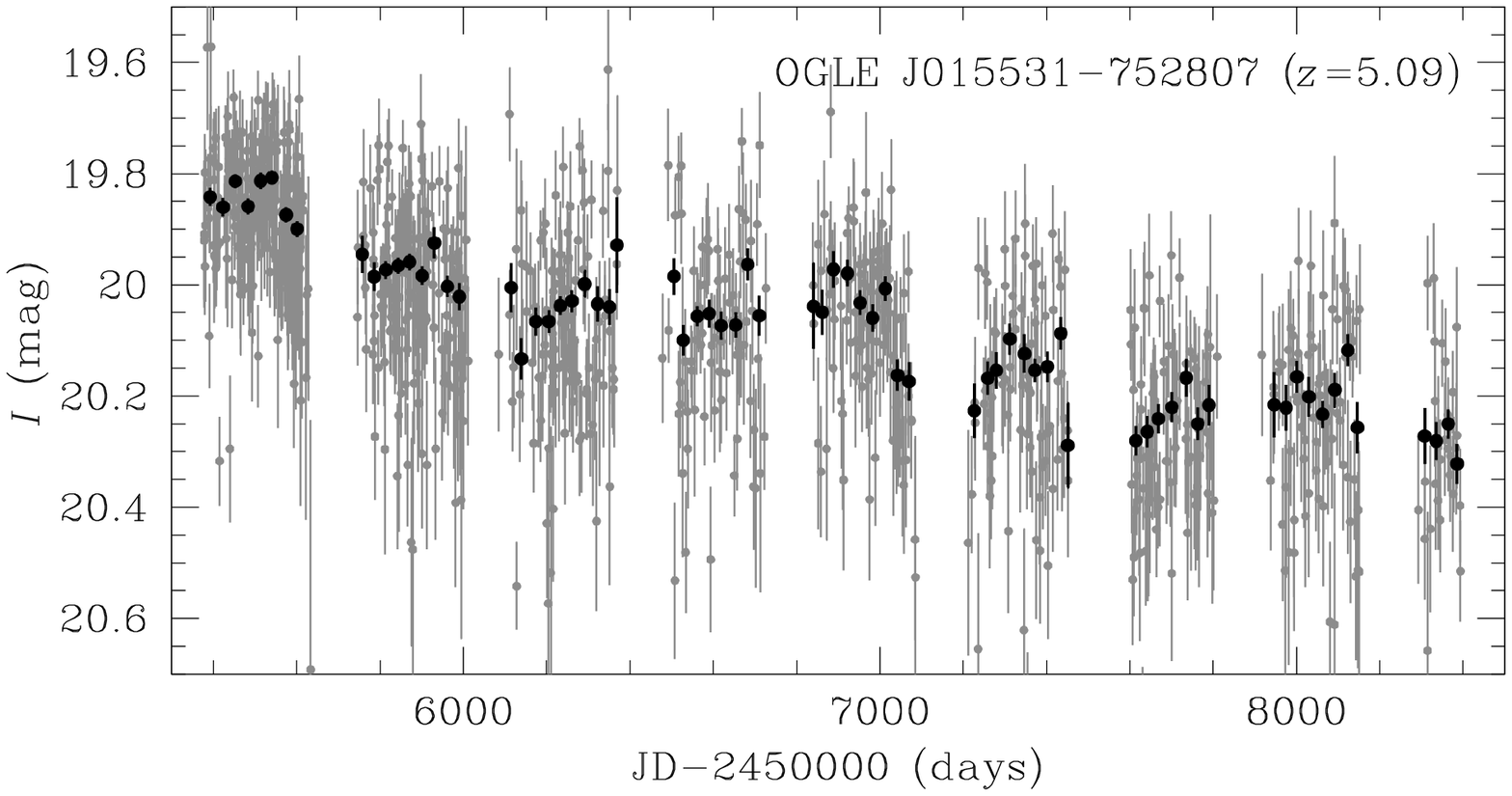}\vspace{0.1cm}
\includegraphics[width=8.2cm]{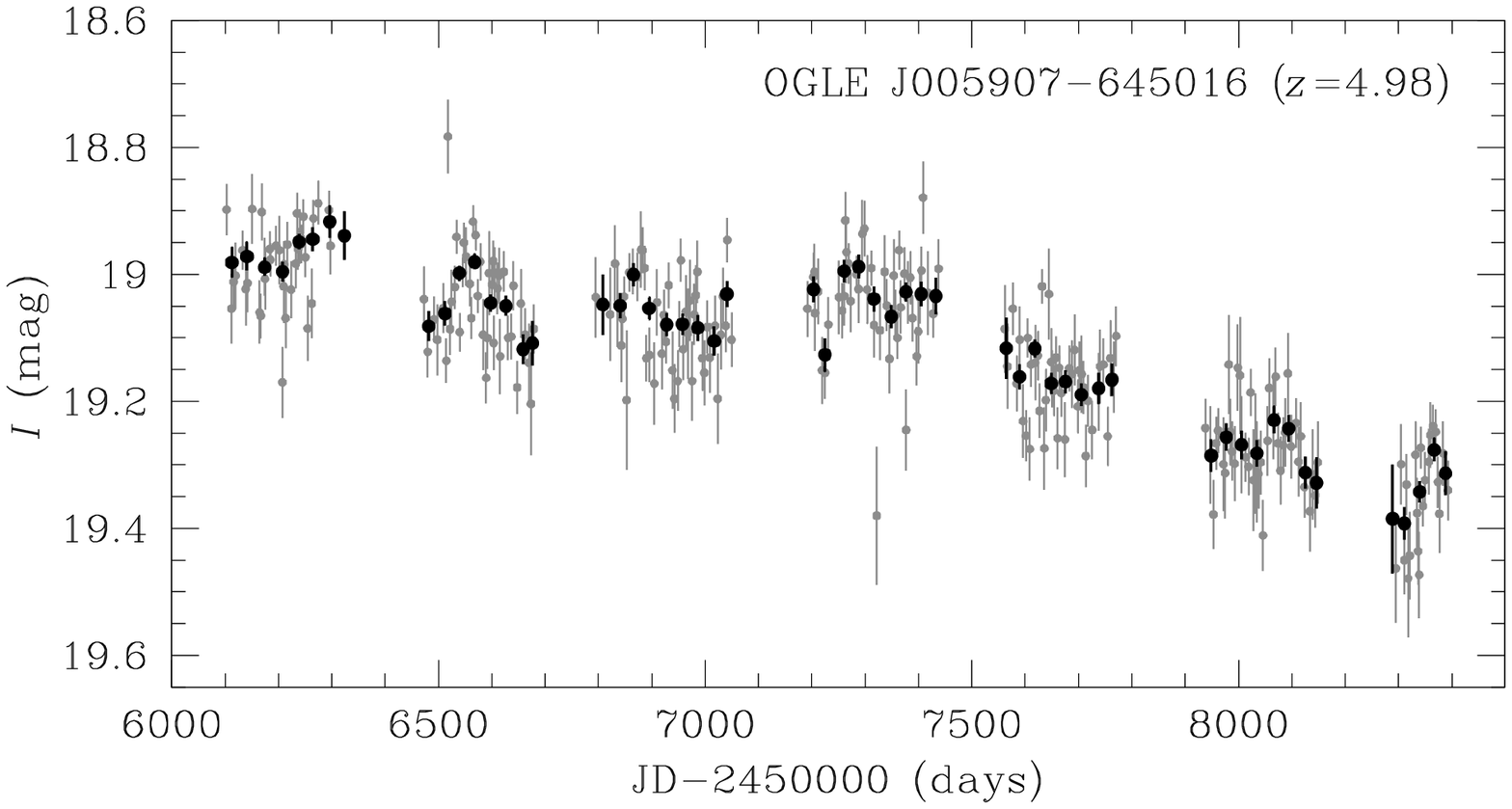}
\caption{The original OGLE-IV $I$-band light curves of \name\ (top; gray points) and \nme2\ (bottom; gray). Both light curves are binned into 30-day-long contiguous bins (black). In both cases, the magnitude change is approximately 0.5 mag. 
The time span is 8.3 (top) and 6.3 (bottom) years, but given time dilation due to the high redshift of our sources, 
in the rest-frame this corresponds to $8.3/(1+5.09)=1.4$ years and $6.3/(1+4.98)=1.1$ years, respectively.}
\label{fig:lc}
\end{figure}

One of the interesting aspects of our search was the presence of AGN variability in the $I$-band. In Figure~\ref{fig:lc} we present the OGLE light curve
for \name. We analyzed the light curve using the structure function (SF) methodology
that measures the typical variability amplitude as a function of time separation between points (see a review in \citealt{2016ApJ...826..118K}). The light curve is short in the rest frame, hence we are not (or weakly) probing the bending SF. We therefore model the SF as a single power law using $SF=SF_0\left(\Delta t/1~{\rm yr}\right)^\gamma$ and obtain the amplitude at 1~year $SF_0=0.39$~mag and the SF slope $\gamma=0.56\pm0.02$, the latter consistent with recent AGN variability studies (e.g., \citealt{2016ApJ...826..118K,2017ApJ...834..111C}).
The measured amplitude of 0.39 mag at rest-frame $\lambda_{\rm emission}\approx 1300$\AA, can be compared to 
typical amplitudes for SDSS AGN in $g$-band at $z=2.7$ and $r$-band at $z=3.7$ (that also probe $\lambda_{\rm emission}\approx 1300$\AA)
to find that it is significantly higher than typical $SF_0 \lesssim 0.15$~mag (from SDSS $g$- and $r$-bands; e.g.,~\citealt{2016ApJ...826..118K}).

We calculated the absolute magnitudes to be $M_I=-27.82$ mag or $M_i=-27.95$ mag using k-corrections\footnote{\url{http://www.astrouw.edu.pl/~simkoz/AGNcalc/}} of $-0.61$~mag and $-0.63$~mag, respectively, and where the monochromatic (at 1350\AA) and bolometric luminosities
are approximately $10^{46.3}$ erg/s and $10^{46.8}$ erg/s (\citealt{2015AcA....65..251K}). 
We also provide the 1450\AA\ rest-frame apparent and absolute magnitudes, 
$m_{\rm 1450}=19.98$ mag and $M_{\rm 1450}=-26.48$ mag, respectively.

From the typical (at lower redshifts) relation between the variability amplitude and the Eddington ratio, we could estimate the \lled\ of about $\sim 0.02$.
This is much lower than what has been found for quasars at higher redshifts, where $0.3<\rled<0.4$ (\citealt{2017ApJ...849...91M,2018arXiv180905584S}).
The variability amplitude--Eddington ratio relation has not been tested at high redshifts. This, in principle, could be done by estimating an independent value of the Eddington ratio from measuring the black hole masses in quasars using the MgII line, but this is beyond the scope of this work, primarily aimed at reporting new high-$z$ quasars.  

\subsection{\nme2}

The second confirmed quasar is \nme2 at $($RA, Decl.$)=($00:59:07.37, $-$64:50:16.11$)$.
The OGLE-IV ID for this source is SMC801.19.759.
The mean extinction corrected $I$-band magnitude is $I=19.11$ mag. The source is not detected
at the $3\sigma$ level in the deep $V$-band, although there is some flux present at this location.
Since this object was not observed with GROND, it has a much scarcer SED coverage. Nevertheless, 
prior to spectroscopic confirmation we obtained the photometric redshift of $z\approx5.0$ for an SED AGN model
that included extinction.

Similarly to the first quasar, the confirmation spectrum for \nme2\ was taken with LDSS3 (Fig.~\ref{fig:spec}, bottom panel). 
This quasar is confirmed in a 900 seconds spectrum taken on 2018 September 29 with 1.0 arcsec seeing and some cirrus clouds.
Again, we used the VPH-RED grism with the 1~arcsec wide ``red'' long-slit. The spectrum shows a Ly$\alpha$ break at around 7310 \AA, 
meaning that it is a quasar at $z\approx5$. The cross-correlation of this spectrum with 
the \cite{2001AJ....122..549V} SDSS AGN spectrum and the \cite{2016A&A...585A..87S} bright QSO spectrum peaks at $r=0.6$ for $z=4.98$ (Figure~\ref{fig:zest}, dashed lines).
The mean magnitude of $I=19.11$ mag at the best estimated redshift of $z=4.98$ translates into
the absolute magnitude $M_I = -28.73$~mag, the monochromatic (at 1350\AA) and bolometric luminosities
of approximately $10^{46.8}$ erg/s and $10^{47.3}$ erg/s, respectively.
The 1450\AA\ rest-frame apparent and absolute magnitudes are
$m_{\rm 1450}=19.72$ mag and $M_{\rm 1450}=-26.70$ mag, respectively.

The SF analysis of the light curve's variability provides the amplitude at 1~year of $SF_0=0.38$~mag and the SF slope $\gamma=0.58\pm0.04$. 
This is yet again a highly variable AGN, although the primary selection indicator for the spectroscopic follow-up was the presence of the
high variability. 

\begin{deluxetable}{lccccc}
\centering
\tabletypesize{\scriptsize}
\tablecaption{Other high-$z$ candidates.\label{tab:other}}
\tablewidth{\columnwidth}
\tablehead{\\ RA & Decl. & $I$     & photo-$z$  & photo-$z$ & var.\\
              &       & (mag) & ext. & no ext. &}
\startdata
23:26:18.01 & $-$71:42:52.70 & 19.19 & 5.3 & 5.3 & no \\ 	
00:15:49.60 & $-$65:23:11.64 & 19.26 & 5.2 & 5.2 & weak \\ 	
00:28:43.69 & $-$79:57:26.20 & 20.00 & 5.2 & 5.2 & weak \\ 	
00:35:25.73 & $-$67:30:44.94 & 19.69 & 5.2 & 5.2 & no \\ 	
00:48:54.66 & $-$66:51:09.11 & 19.12 & 5.3 & 5.3 & no \\ 	
00:56:27.70 & $-$63:47:41.10 & 19.53 & 5.0 & 5.0 & no \\ 	
01:02:20.09 & $-$65:03:01.30 & 19.11 & 5.3 & 5.2 & yes \\ 	
01:21:19.23 & $-$66:23:03.11 & 19.75 & 5.0 & 5.0 & weak \\ 	
01:31:19.13 & $-$75:48:39.01 & 20.27 & 5.2 & 5.0 & no \\ 	
01:32:00.72 & $-$64:25:03.89 & 19.56 & 5.2 & 5.2 & no \\ 	
01:48:23.94 & $-$71:27:47.39 & 19.06 & 4.7 & 4.7 & weak \\ 	
02:12:25.34 & $-$70:47:35.42 & 19.51 & 5.2 & 5.2 & no \\ 	
03:16:35.75 & $-$73:39:06.31 & 19.65 & 5.3 & 5.0 & no \\ 	
03:26:40.47 & $-$72:07:08.10 & 19.94 & 5.0 & 5.0 & no \\ 	
03:52:06.25 & $-$79:32:09.25 & 19.35 & 5.2 & 5.2 & no \\ 	
04:05:13.44 & $-$71:57:13.06 & 19.63 & 5.0 & 5.2 & no \\ 	
04:19:02.55 & $-$73:12:13.76 & 18.92 & 5.0 & 5.0 & no \\ 	
04:26:32.77 & $-$76:59:00.26 & 19.02 & 5.2 & 5.2 & no \\ 	
04:28:59.91 & $-$74:22:41.26 & 19.67 & 5.2 & 5.2 & no \\ 	
04:37:22.76 & $-$79:51:08.93 & 19.56 & 5.0 & 5.0 & weak \\ 	
04:51:33.80 & $-$63:22:06.90 & 19.80 & 5.0 & 5.0 & no \\ 	
04:58:23.78 & $-$76:51:40.63 & 19.68 & 5.2 & 5.7 & weak \\ 	
05:04:43.39 & $-$76:11:45.18 & 19.52 & 4.7 & 4.7 & no \\ 	
05:11:47.47 & $-$58:12:52.40 & 19.34 & 5.0 & 5.2 & weak \\ 	
05:17:14.65 & $-$78:03:32.99 & 19.54 & 5.0 & 5.2 & weak \\ 	
05:28:17.80 & $-$54:07:19.23 & 19.56 & 5.0 & 5.0 & weak \\ 	
05:30:27.69 & $-$61:49:46.00 & 20.29 & 5.3 & 5.8 & no \\ 	
05:41:05.62 & $-$61:38:49.37 & 19.78 & 5.2 & 5.2 & no \\ 	
05:45:54.93 & $-$75:38:47.04 & 19.74 & 5.3 & 5.2 & no \\ 	
06:03:31.77 & $-$61:50:29.98 & 19.67 & 5.3 & 5.2 & yes \\ 	
06:52:55.15 & $-$70:53:28.19 & 20.02 & 5.2 & 5.0 & weak \\ 	
\enddata
\tablecomments{The header reads: ``photo-$z$ ext.'' means the photometric redshift estimate with the extinction included in the fit, while
``photo-$z$ no ext.'' means the photometric redshift estimate excluding the extinction from the fit (fixed to $A_V=0.0$ mag), 
``var./remarks'' informs if the source appears to be variable in the OGLE data (yes/weak/no).}
\end{deluxetable}

\section{Other candidates}
\label{sec:other}

In Table~\ref{tab:other}, we present basic properties of the remaining candidates selected with the method presented in Section~\ref{sec:method}. 
All sources were checked against proper motions in OGLE (one object rejected), and all of them were fit with an SED that includes or excludes the internal AGN extinction. By definition (Section~\ref{sec:method}) and as a minimum requirement, all the sources have constraints in the $V$- and $I$-bands, and the WISE mid-IR bands. 

We have cross-matched our list of candidates against the {\it Gaia} Data Release 2 catalogue (\citealt{2018A&A...616A...1G,2018A&A...616A...2L}) to find 20 matches of which 12 have proper astrometric solutions (objects with $G<20.9$ mag). There are four GAIA sources with well measured parallaxes ($<20$\%) and relatively high proper motions of 5--28 mas/yr that very likely can not be high-$z$ quasars. We remove them from our sample and the remaining sample consists of \cand\ good high-$z$ quasar candidates.

None of the candidates including the two confirmed quasars is present in the Simbad database (with a large matching radius of $r<30$ arcsec).
There is no candidate or quasar present in the Galex database with the radius of $r<5$ arcsec (\citealt{2011Ap&SS.335..161B}) nor
X-ray detection as reported in Vizier (\citealt{2000A&AS..143...23O}). The closest match to our objects in the NED database is at 9 arcsec, hence
none of our candidates and quasars has a counterpart in that database.


\section{Summary}
\label{sec:summary}

In this paper we presented a high-$z$ quasar selection method designed for the OGLE survey that includes a combination of the   
$V$-band dropout method, the optical--mid-IR selection method, and AGN variability. By matching the deep OGLE data and WISE data 
for the Magellanic Clouds, we were able to select \cnd\ quasar candidates. 

The most variable candidate was subsequently observed with GROND
in $g'r'i'z'JHK$ filters and together with archival data they allowed us to construct an SED from 0.4 to 24 $\mu$m at 16 wavelengths. The best AGN SED fit 
suggested $z\approx 5.2$ and $A_V\approx 0.28\pm0.02$ mag. We observed spectroscopically this candidate with the 6.5m Magellan Telescope and
confirmed it as a genuine quasar at a redshift of $z=5.09$. This is the most distant (and variable) object identified in the OGLE survey to date.
The monochromatic luminosity of \name\ at 1350 \AA\ is $10^{46.3}$ erg/s, while the bolometric luminosity is $10^{46.8}$ erg/s. 

Our second confirmed quasar, \nme2, was not observed with GROND, but given the OGLE and WISE data, we estimated the redshift to be $z\approx5.0$.
A subsequent spectroscopic follow-up with the 6.5m Magellan Telescope provided the redshift of $z=4.98$.

Both sources seem to be significantly more variable (0.38--0.39 mag) at this rest-frame wavelength ($\sim$1300\AA) than quasars at lower redshifts 
(but at the same rest-frame wavelength) having typical variability amplitude at one year rest-frame of $<0.15$ mag (e.g., \citealt{2016ApJ...826..118K}). 

Having a sample consisting of two objects it is virtually impossible
to place any constraints on the AGN variability evolution with time. 
Our two quasars are outstanding because sources when the Universe was 1.7 Gyr old 
(at $z\approx3.7$) show significantly smaller variability amplitudes than our quasars observed 1.2 Gyr after Big Bang (half a Gyr time difference).
Sources with extreme properties, however, are not uncommon in lower-$z$ AGN samples. Therefore, this variability issue 
can be resolved by studying variability of a larger AGN sample at $z\approx5$. This is entirely possible now, because we also provide a list of \cand\ AGN candidates at $z\approx5$, where all of them have nearly a decade long light curves from OGLE. 
A sample of high-$z$ AGN with a decade-long light curves will be also increased by the Large Synoptic Survey Telescope (LSST; e.g., \citealt{2008arXiv0805.2366I,2011ApJ...728...26M}), providing light curves for sources 3--4 mag fainter than OGLE.


\acknowledgments

SK acknowledges the financial support of the Polish National Science Center through the
OPUS grant number 2014/15/B/ST9/00093. OGLE is supported from the MAESTRO grant number 2014/14/A/ST9/00121 to AU.
LW and MG are supported by the Polish National Science Centre grant OPUS 2015/17/B/ST9/03167.
APJ is supported by NASA through Hubble Fellowship grant HST-HF2-51393.001 awarded by the Space Telescope Science Institute, which is operated by the Association of Universities for Research in Astronomy, Inc., for NASA, under contract NAS5-26555.
Part of the funding for GROND (both hardware as well as personnel)
was generously granted from the Leibniz-Prize to Prof. G. Hasinger
(DFG grant HA 1850/28-1). We acknowledge the support of Markus Rabus
for the GROND observations.

This paper includes data gathered with the 6.5 m Magellan Telescopes located at Las Campanas Observatory, Chile.
This publication makes use of data products from the Wide-field Infrared Survey Explorer, which is a joint project of the University of California, Los Angeles, and the Jet Propulsion Laboratory/California Institute of Technology, and NEOWISE, which is a project of the Jet Propulsion Laboratory/California Institute of Technology. WISE and NEOWISE are funded by the National Aeronautics and Space Administration.
This research has made use of the NASA/IPAC Extragalactic Database (NED), which is operated by the Jet Propulsion Laboratory, California Institute of Technology, under contract with the National Aeronautics and Space Administration.

\facilities{Magellan: Clay (LDSS3 spectrograph); Max Planck: 2.2m (GROND); Warsaw Telescope: 1.3m (OGLE)}


\end{document}